\documentclass[showkeys,
 reprint,
 amsmath,amssymb,
 aps,
]{revtex4-2}

\usepackage{xcolor}
\usepackage{textcmds}
\usepackage{graphicx}
\usepackage{subcaption}
\usepackage{dcolumn}
\usepackage{bm}
\usepackage{overpic}
\usepackage{multirow}

\begin{document}

\title{Reducing TLS Losses in Superconducting Circuits Using Solid Neon}

\author{A. Palmer$^{1,2}$}
    \email{andrew.r.palmer@student.unsw.edu.au}%
\author{K. Matkovic$^1$}
\author{P. Russell$^{1,2}$}
\author{E. Helgemo$^{1,2}$}
\author{L. Delventhal$^{1,2}$}
\author{Y. Ashlea Alava$^{1,2}$}
\author{K. Zuo$^3$}
\author{S.N. Coppersmith$^{1,2}$}%
\author{M.C. Cassidy$^{1,2}$}%
 \email{maja.cassidy@unsw.edu.au}
\affiliation{%
 $^{1}$ ARC Centre of Excellence for Quantum Computer Performance and Integration, University of New South Wales, Sydney, 2052, NSW, Australia
}%
\affiliation{%
 $^{2}$ School of Physics, University of New South Wales, Sydney, 2052, NSW, Australia
}%
\affiliation{
 $^{3}$ School of Physics, University of Sydney, Sydney, 2006, NSW, Australia
}%
\hspace{10pt}
\date{\today}

\begin{abstract}
Two level systems (TLSs) are a major source of loss in superconducting devices, limiting the performance of qubits, detectors, and resonators. These defects arise from a range of sources, including surface oxides, crystal defects, as well as fabrication residues which are often difficult to remove. Using NbTiN coplanar waveguide resonators, we show that depositing and subsequently removing a solid neon film can significantly improve device performance. Analysis of the temperature and power dependence indicates that this enhancement arises from reduced TLS dissipation, which we attribute to the redistribution and removal of surface contaminants during neon deposition and evaporation. Atomic force microscopy images of contaminated hexagonal boron nitride flakes before and after neon exposure show that neon treatment can redistribute and remove surface residues. These results demonstrate a simple, non-destructive approach for mitigating TLS losses in superconducting devices and other sensitive materials that cannot withstand aggressive cleaning processes.
\end{abstract}

\maketitle{}


\section{\label{sec:level1}INTRODUCTION}
Coplanar waveguide (CPW) resonators are a fundamental component of many superconducting devices and circuits, including  single-photon detectors \cite{dayBroadbandSuperconductingDetector2003}, quantum limited amplifiers \cite{castellanos-beltranWidelyTunableParametric2007, hoeomWidebandLownoiseSuperconducting2012} and quantum processors \cite{manucharyanFluxoniumSingleCooperPair2009, wallraffStrongCouplingSingle2004}. Their performance is often limited by energy dissipation \cite{mcraeMaterialsLossMeasurements2020} which broadens the linewidth and reduces the internal quality factor $Q_i$. Excitations of environmental two-level systems (TLSs) are widely recognised as a dominant sources of loss in many supercondcting circuits \cite{mullerUnderstandingTwolevelsystemsAmorphous2019}, especially at millikelvin temperatures and low excitation powers. Considerable efforts have been focussed at identifying the sources of these losses and developing ways to reduce them \cite{crowleyDisentanglingLossesTantalum2023, foshatQuasiparticleDynamicsNbN2025, sageStudyLossSuperconducting2011, tominagaEnhancingIntrinsicQuality2025, lisenfeldElectricFieldSpectroscopy2019, lescanneExponentialSuppressionBitflips2020}.\\

\begin{figure*}[t]
    \centering
    \includegraphics{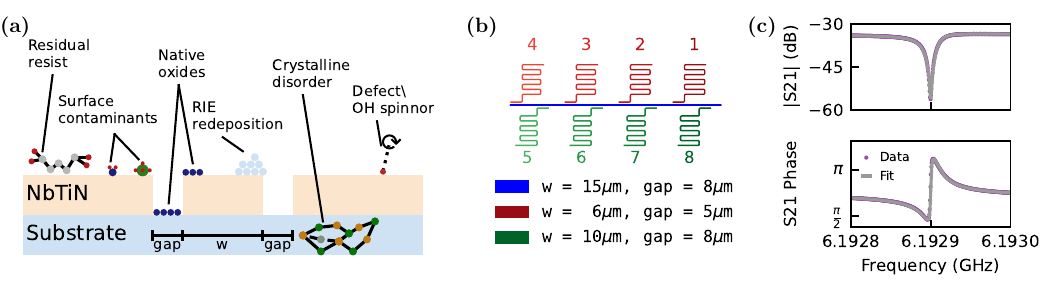}
    \captionsetup{justification=raggedright, singlelinecheck=false}
    \caption{\label{fig:figure1} \textbf{Sources of TLS losses and device specifications.} %
    \textbf{(a)} Illustration of device material layers (not to scale) and possible sources of TLS losses. Key resonator dimensions, the central conductor width (w) and gap to the ground plane are indicated.%
    \textbf{(b)} Design schematic of the devices consisting of eight numbered resonators coupled to a single feedline (blue), resonators 1-4 (red) have a narrow structure design while resonators 5-8 (green) have a wider structure design. %
    \textbf{(c)} Amplitude and phase of the forward transmission coefficient ($S_{\text{21}}$) as a function fo frequency for a single resonator (resonator 6). The solid lines indicate the fit to the transmission model from which the internal ($Q_{i}$) and coupling ($Q_c$) quality factors have been extracted.}
\end{figure*}

Although the exact microscopic origin of many TLSs remains uncertain \cite{mullerUnderstandingTwolevelsystemsAmorphous2019}, they are generally associated with atomic or electronic configurations that can occupy two metastable states. A TLS can absorb energy from the resonator's electric field and subsequently relax, transferring energy to the environment, for example via emission of a phonon \cite{mcraeMaterialsLossMeasurements2020}. The resulting dissipation depends on temperature and excitation power when TLSs become saturated as their populations and transition rates change. Some known sources of TLSs include crystal defects within the substrate material \cite{lisenfeldElectricFieldSpectroscopy2019}, native oxides on the surface of the substrate or superconductor \cite{lozanoLowlossAtantalumCoplanar2024}, chemical adsorbates  \cite{degraafSuppressionLowfrequencyCharge2018}, and contaminants from the fabrication process such as residual resist polymers \cite{quintanaCharacterizationReductionMicrofabricationinduced2014} or etch byproducts \cite{sandbergEtchInducedMicrowave2012}, as illustrated in Figure \ref{fig:figure1}a. Changing the device geometry can reduce the electric field participation ratio of these lossy interfaces \cite{tominagaEnhancingIntrinsicQuality2025}, however, it does not remove the underlying defects. Because TLSs possess an enormous number of degrees of freedom, they are able to contribute to loss over a wide range of resonator frequencies. 
Several approaches have been used to reduce TLS loss. This includes oxide removal by chemical treatments \cite{lozanoLowlossAtantalumCoplanar2024}, oxide-growth suppression through encapulation or passivation  \cite{balSystematicImprovementsTransmon2024, guptaHighTemporalStability2026}, absorbate removal through high temperature annealing \cite{degraafSuppressionLowfrequencyCharge2018} or ultraviolet light exposure \cite{mergenthalerUltrahighVacuumPackaging2021}, and removal of fabrication residues using strong acids \cite{crowleyDisentanglingLossesTantalum2023}. These methods can be effective, but some are chemically aggressive, require elevated temperatures, or are unsuitable for materials that are sensitive to processing. A treatment compatible with a wider range of superconductors and two-dimensional materials would therefore be useful.\\

Cryogenic noble-gas deposition offers a possible non-reactive way to modify surface contaminants. Previous experiments coating superconducting circuits with noble gases at cryogenic temperatures have observed increases in resonator quality factors \cite{tianNbTiNNanowireResonators2025, matkovicCharacterizingNeonThin2025, valentiHydrogenCrystalsReduce2024}. The persistence of those changes after the noble gas is removed, and the mechanism responsible for them, remain unclear. Noble-gas films have also been proposed to affect radiation-induced quasiparticle loss \cite{valentiHydrogenCrystalsReduce2024} and device thermalisation \cite{lucasQuantumBathSuppression2023}.\\

Here we study NbTiN CPW resonators before, during and after cryogenic neon deposition.  By comparing measurements from multiple cooldowns before and after the neon treatment, we observe an increase in the internal quality factor, ($Q_i$), that persists even after the neon has been removed. Fitting the power- and temperature-dependent losses to an analytical model indicates that the improvement arises from a reduction in TLS dissipation caused by the redistribution or removal of fabrication residues from the device surface, rather than changes in dielectric participation, cooling of the TLS bath, or shielding from quasiparticle-generating radiation. The observed increase in $Q_i$ was then replicated on an identical chip in a different dilution refrigerator setup. To investigate the underlying mechanism, we exposed a contaminated hexagonal boron nitride flake to the same neon treatment. Atomic force microscopy images acquired before and after treatment reveal a redistribution of surface contaminants, supporting the hypothesis that the neon process removes or displaces fabrication residues responsible for excess TLS loss.

\begin{figure*}[t]
    \centering
    \includegraphics{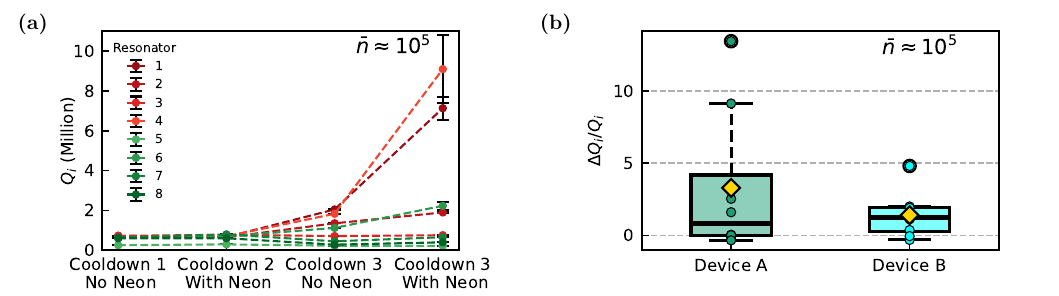}
    \captionsetup{justification=raggedright, singlelinecheck=false}
    \caption{\label{fig:figure2} \textbf{Evolution of $\mathbf{Q_i}$ at $\mathbf{20~mK}$}. %
    \textbf{(a)} Change in $Q_i$ over different cooldowns applying and removing neon for each resonator, the dashed line connects points for each resonator for visual clarity. %
    \textbf{(b)} Change in $Q_i$ between initial and final measurement for both devices, each dot represents the change in $Q_i$ for an individual resonator, with the gold diamond indicating the mean improvement.}
    
\end{figure*}

\section{\label{sec:level2}SETUP AND ANALYSIS METHODS}

The measured devices were fabricated from NbTiN films deposited on high-resistivity silicon substrates. We observed similar behaviour across two resonator samples measured in different setups. Both devices contain eight quarter wavelength hanger resonators multiplexed to a single feedline, as shown in Figure \ref{fig:figure1}b. In the main text of the paper we focus on the results from device A, with additional data from device B contained in the Supplementary Materials.  In device A, resonators $1$-$4$, (red), have frequencies spaced between $5.75$ and $6.06$ GHz and have a central conductor width of $6~\mu m$ with a gap of $5~\mu m$, while resonators $5$-$8$, (green), have frequencies spaced between $6.10$ and $6.49$ GHz and have a central conductor width (w) of $10~\mu m$ with a gap of $8~\mu m$. The resonators are capacitively coupled to the feedline via a $200~\mu m$ length of superconductor, with a separation distance of $35~\mu m$. The sample was bonded to a PCB and placed in a hermetically sealed copper cell which was attached to the mixing chamber of a Bluefors dilution refrigerator with a base temperature of $\sim20~mK$. Full details of the resonator design and fabrication are provided in the Supplementary Materials \ref{sect:app_resdesign} and \ref{sect:app_sampleprep}. \\

The resonators were characterised by performing transmission spectroscopy using a commercial Vector Network Analyzer (VNA) with the output power attenuated to control the average number of photons in the resonator, $\bar{n}$. Fig. \ref{fig:figure1}c shows an example of the amplitude and phase of the transmission coefficient ($S_{\text{21}}$) for a single resonator. The device was initially characterised at $20~mK$ (cooldown 1) and then warmed up to room temperature. In the second cooldown, the sample was measured at $20~mK$, and then characterised after neon deposition.  Finally, the sample underwent a third cooldown where it was characterised both before and after neon deposition. Full details on the measurement setup and timeline are provided in Supplementary Materials \ref{sect:app_measurement_setup}.\\

The neon deposition process used is similar to described in other works \cite{zhouSingleElectronsSolid2022, matkovicCharacterizingNeonThin2025}. To coat the sample with neon, all plates on the cryostat were heated to $26~K$, above the triple point temperature of neon ($24.556~K$, \cite{preston-thomasInternationalTemperatureScale1990}), before a set amount of room temperature gas was released into the cell via a capillary feedline. The neon gas condenses into liquid on the sample, where it crystallises once the temperature was reduced below the triple point. Full details of the neon deposition process are provided in Supplementary Materials \ref{sect:app_neondep}. The presence of neon on the chip was confirmed by a shift of the resonance frequency, $f_r = \frac{1}{2\pi\sqrt{LC}}$ \cite{pozarMicrowaveEngineering2012}, due to increased capacitance of the resonator, $C$, from the dielectric neon ($\epsilon = 1.244$ \cite{inuiOnChipLevitatedNeon2025}). \\

The amplitude and phase response for each measurement was fit using \cite{probstEfficientRobustAnalysis2015}

\begin{equation}  \label{eq:1}
    S_{21}(f) = A e^{i \alpha}e^{-2\pi i f \tau} \left( 1 - \frac{\frac{Q_l}{|Q_c|}e^{i\phi}}{1 + 2iQ_l\frac{f - f_r}{f_r}}\right).
\end{equation}

\begin{figure*}[t ]
    \centering
    \includegraphics{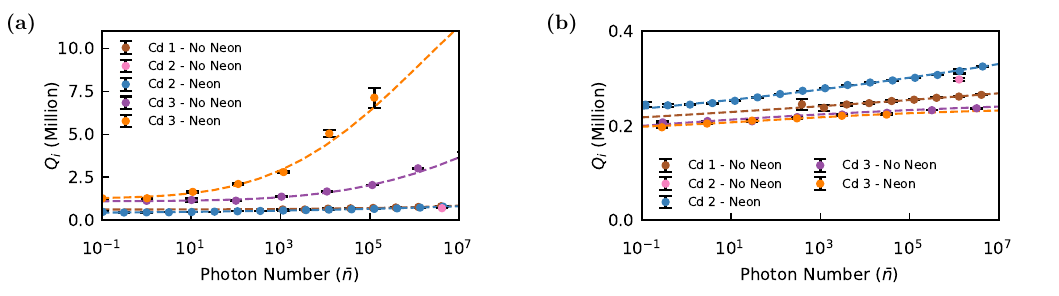}
    \captionsetup{justification=raggedright, singlelinecheck=false}
    \caption{\label{fig:figure3} \textbf{Power dependence of $\mathbf{Q_i}$ at $\mathbf{20~mK}$}. %
    $Q_i$ as a function of power for \textbf{(a)}  resonator 1, that showed a significant improvement in $Q_i$, and \textbf{(b)} resonator 5, representative of one that showed no increase in $Q_i$ over the measurement sequence.
    The dashed lines shows the fit of $Q_i$ from equation (\ref{eq:2}) for each cooldown.}\end{figure*}

Here, $A$ and $\alpha$ are the baseline transmission amplitude and phase, $f$ is the probe frequency, $\tau$ is the phase adjustment for the signal travel time, $Q_l$ is the total, or loaded, quality factor and $Q_c$ is the coupling quality factor. The phase shift due to impedance mismatches in the measurement circuit is captured by $\phi$. From this fit we can extract $f_r$, $Q_c$ and $Q_i$ through the relationship that $\frac{1}{Q_i} = \frac{1}{Q_l} - \frac{1}{Q_c}$.\\

We further parametrize $Q_i$ using a model that represents the total loss as the sum of TLS-associated, quasiparticle associated and other contributions \cite{crowleyDisentanglingLossesTantalum2023, moshelMaterialLossModel2025}, 

\begin{equation}  \label{eq:2}
    \frac{1}{Q_i} = \frac{1}{Q_{TLS}(\bar{n}, T)} + \frac{1}{Q_{QP}(T)} + \frac{1}{Q_{other}}.
\end{equation}

Here, $Q_{TLS}$ describes the power and temperature dependent TLS contribution, $Q_{QP}$ describes the temperature-dependent quasiparticle contribution, and $Q_{other}$ captures all other losses which are not dependent on temperature and power.\\

At $20~mK$, we assume the losses from quasiparticles are negligible so that $\frac{1}{Q_i} \approx \frac{1}{Q_{TLS}(\bar{n}, T)} + \frac{1}{Q_{other}}$. The TLS losses can be parameterised by the model \cite{crowleyDisentanglingLossesTantalum2023, mcraeMaterialsLossMeasurements2020, moshelMaterialLossModel2025}

\begin{equation} \label{eq:3}
    Q_{TLS}(\bar{n}, T) = Q_{TLS, 0} \frac{\sqrt{1 + \frac{\bar{n}}{n_c}}}{\tanh\left(\frac{\hbar \omega}{2k_BT}\right)}.
\end{equation}

$Q_{TLS, 0}$ sets the low power, low temperature limit of TLS losses, which captures both the density of TLSs and how strongly they couple with the resonator, known as the filling factor. The number of photons needed to saturate the TLS systems, $n_c \propto 1/T_1 T_2$ is inversely proportional to the product of the TLS relaxation ($T_1$) and dephasing ($T_2$) times, these terms describe the rate at which the energy from TLSs are dissipated to the environment. We further approximate that $T_1 \propto \tanh\left(\frac{\hbar \omega}{2k_BT}\right)$, $(T_2)^{-1} \propto k_BT^{\beta_1}$  together fit parameter $\beta_2$ to account for the non linear response of different TLS modes to increasing number of photons \cite{crowleyDisentanglingLossesTantalum2023}. \\





To model how the $Q$ factor responds to temperature, we need to account for quasiparticle losses which dominate at higher temperatures. Quasiparticle losses can be parameterised by \cite{crowleyDisentanglingLossesTantalum2023, moshelMaterialLossModel2025}

\begin{equation} \label{eq:5}
    Q_{QP}(T) = Q_{QP, 0} \frac{e^{\frac{\Delta}{k_B T}}}{\sinh\left(\frac{\hbar \omega}{2k_BT}\right) K_0 \left(\frac{\hbar \omega}{2k_BT}\right)}.
\end{equation}

Here $Q_{QP_0}$ is the total quasiparticle number scaling parameter,  $\Delta$ is the superconducting gap, approximated as $\Delta \approx 1.764k_BT_c$ \cite{crowleyDisentanglingLossesTantalum2023}, where $T_c$ is the critical temperature of the superconductor, and $K_0$ is the zeroth order modified Bessel function of the second kind. Further details of the fitting procedure are contained in the Supplementary Materials \ref{sect:app_fits}. 

\section{\label{sec:level3}$Q_i$ IMPROVEMENT AFTER NEON DEPOSITION AND REMOVAL}

Figure \ref{fig:figure2}a shows the evolution of $Q_i$ across the different cooldowns at $20~mK$ and high power ($\bar{n}  \approx 10^5$) in device A. We observe a similar $Q_i$ value between the initial characterisation and after neon was first deposited in cooldown 2. However, once the neon was removed from the sample by warming to room temperature, and cooled down a third time, $Q_i$ increased in resonators 1, 2, 4 and 6, with the others staying constant, or decreasing slightly. For those resonators that showed an improved $Q_i$, this improvement was further enhanced by an additional neon deposition. Figure \ref{fig:figure2}b summarizes the change in $Q_i$ for each resonator across the two samples between the initial characterisation and the final measurement. On average, resonators in device A improved to $4.3$ times their initial $Q_i$, while those in device B improved to $2.4$ times their initial $Q_i$.

\section{\label{sec:level4}POWER AND TEMPERATURE DEPENDENCE}

To isolate the different loss mechanisms, we measured the temperature dependence of  $Q_i$ across a range of powers. Figure \ref{fig:figure3} compares $Q_i$ as a function of photon number at $20~mK$ for resonator 1, which showed a significant improvement after neon removal, and resonator 5, which did not. Fits to the model described in Section \ref{sec:level2} are shown as dashed lines in Figures \ref{fig:figure3}a and \ref{fig:figure3}b. For resonator 1 (Figure \ref{fig:figure3}a), $Q_i$ increases across the measured power range and does not saturate. While the improvement is most evident at higher powers, in the single photon regime (n=1), which is most applicable to quantum computing applications, $Q_i$ increased  from 
 $4.7 \times 10^5$ to $1.4 \times 10^6$, an increase of $\sim190\%$. In comparison, for resonator 5 (Figure \ref{fig:figure3}b) $Q_i$ first slightly increases with the first neon deposition, and then decreases for the subsequent removal and neon deposition.  Results for the other resonators are shown in Supplementary Materials \ref{sect:app_fits} and \ref{sect:app_device2}. 
\\

\begin{figure*}[t]
    \centering
    \includegraphics{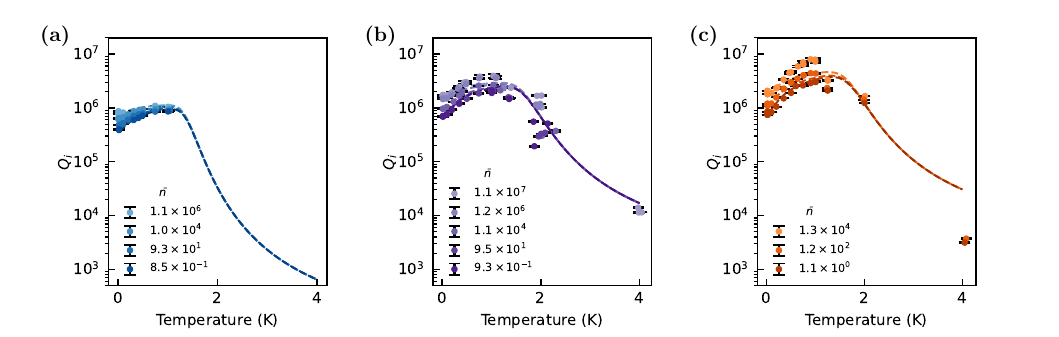}
    \captionsetup{justification=raggedright, singlelinecheck=false}
    \caption{\label{fig:figure4} \textbf{Fit of $Q_i$ across a range of temperatures.}  Observed temperature dependence of $Q_i$ along with the fitted model for Resonator 6 in cooldown 2 with neon \textbf{(a)}, cooldown 3 without neon \textbf{(b)} and cooldown 3 with neon  \textbf{(c)}.}
\end{figure*}

Figure \ref{fig:figure4} shows the temperature and power responses of resonator $6$ as neon is deposited, removed and redeposited on the sample. During the first neon deposition (Figure \ref{fig:figure4}a), $Q_i$  shows a weak dependence on increasing both microwave power and temperature, with smaller increases observed at high power. After the neon is removed (Figure \ref{fig:figure4}b), and redeposited (Figure \ref{fig:figure4}c), the  temperature response of $Q_i$ becomes stronger across the full range of microwave power. In contrast, resonators that showed no improvement in $Q_i$ retained a flat temperature dependence, similar to Figure \ref{fig:figure4}a (see Supplementary Information). \\

Beyond 1 K, $Q_i$ decreases in all resonators as quasiparticle related losses dominate. Previous studies have found that a thin layer of hydrogen can shield superconducting resonators from incident radiation, reducing the contribution of quasiparticles to the loss \cite{valentiHydrogenCrystalsReduce2024}. By comparing $Q_i$ with and without neon at $4~K$, where thermal quasiparticle dissipation dominates, we  conclude that neon does not significantly reduce the quasiparticle losses in the resonator by shielding the device from high energy radiation (see Supplementary Materials \ref{sect:app_quasiparticles}).

\section{\label{sec:level5}ATTRIBUTION OF LOSSES}

We fit the combined power and temperature dependent data to the model described in Section \ref{sec:level2} to extract the TLS dependent losses. \\

\begin{figure*}
    \centering
    \includegraphics{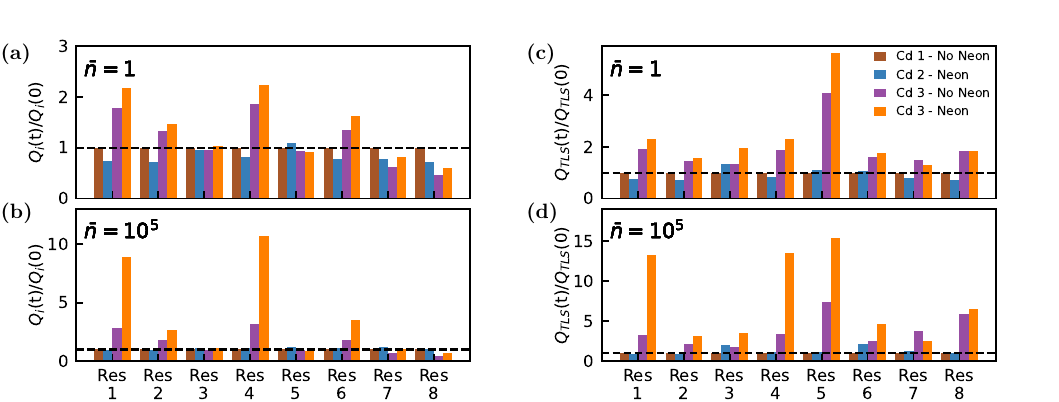}
    \captionsetup{justification=raggedright, singlelinecheck=false}
    \caption{\label{fig:figure5} \textbf{$\mathbf{Q_{TLS}}$ attribution}  Change in $Q_{i}$ for each resonator relative to first measurement at \textbf{(a)} $\bar{n} = 1$ and \textbf{(b)} $\bar{n} = 10^5$.  Change in $Q_{TLS}$ for each resonator relative to first measurement at \textbf{(c)} $\bar{n} = 1$ and \textbf{(d)} $\bar{n} = 10^5$.}
\end{figure*}

Figure \ref{fig:figure5} shows the relative change in fitted $Q_i$ and $Q_{TLS}$ compared to the initial characterisation for all cooldowns and resonators for both the single photon ($\bar{n} = 1$) and high power ($\bar{n} = 10^5$) regimes. While the first application of neon decreases $Q_{TLS}$ for almost all resonators, removing the neon increases $Q_{TLS}$ well beyond the original $Q_{TLS}$ of the sample. Further addition of neon continues to increase $Q_{TLS}$, up to a factor of 6 times in the single photon regime and more than an order of magnitude at high power.  Interestingly, an increase in the fitted $Q_{TLS}$ is observed even for resonators that did not show an increase in overall $Q_i$. This indicates a reduction in the $Q_{Other}$ parameter between cooldowns. This decrease in $Q_{other}$ could be explained by a reduction in the $T_1$ or $T_2$ times of the coupled TLSs. When the lifetimes are sufficiently short, the $\bar{n}/n_c$ term in equation (\ref{eq:3}) becomes vanishingly small, resulting in a weak power and temperature dependence that causes their loss contribution to be absorbed into the $Q_{other}$ parameter, rather than $Q_{TLS}$. The non-uniform response shows multiple dissipation mechanisms exist at the same time and suggests neon processing can locally redistribute the surface contaminants rather than uniformly reduce all losses.\\

\begin{figure*}
    \centering
    \includegraphics{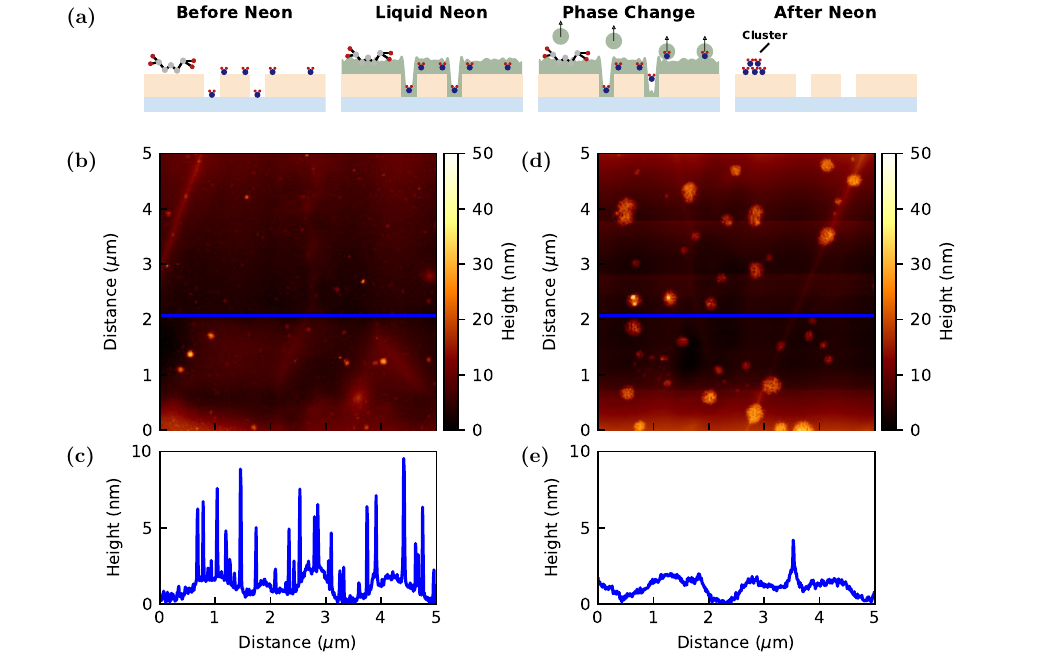}
    \captionsetup{justification=raggedright, singlelinecheck=false}
    \caption{\label{fig:figure6} \textbf{Surface cleaning} \textbf{(a)} Illustration of proposed cleaning mechanism. \textbf{(b)} AFM image of HBN flake contaminated with polymer particles before neon treatment. \textbf{(c)} Data slice of height shown by blue line before neon treatment. \textbf{(d)} AFM image of HBN flake after neon treatment, where small polymer contaminants from before have been moved into larger clusters. \textbf{(e)} Data slice of height shown by blue line after neon treatment. }
\end{figure*}

Given that $Q_{i}$ and $Q_{TLS}$ improve even after the neon is removed from the device and has been thermally cycled, we  conclude that the results are not from changes to the energy participation ratio due to the neon filling the gaps between the resonator central conductor and ground plane. A plausible explanation is that the increase in $Q_{TLS}$ results from the neon removing or redistributing contaminants on the surface of the chip, as illustrated in Figure \ref{fig:figure6}a. We hypothesize that in the first neon deposition, fabrication residues on the top of the superconductor are pushed by the neon into the active area of the resonator, increasing the surface participation ratio of these molecules. When the sample is brought to room temperature and the neon evaporates, these contaminants are taken with it. As no systematic trend is seen for resonators with wide gaps (resonators 1-4) compared to narrow gaps (resonators 5-8), the dominant contribution of TLS is likely from those residing at the surface of the supperconductor, rather than the substrate.\\

To support this hypothesis, we expose a flake of hexagonal boron nitride (hBN) to frozen neon. While the hBN surface differs from the CPW surface in topology and chemistry, this provides qualitative insights into the effect of neon on surface contaminants. This flake is contaminated with polymer residues from the exfoliation process. Figure \ref{fig:figure6}b--e shows atomic force microscopy (AFM) images taken before and after the treatment which show that the polymer residues are moved over micrometer scale distances into clusters, leaving large patches of the surface free of these contaminants. This clumping of residual particles may explain the observed results as the number of TLSs on the surface of the chip have been reduced, or, moved such that coupling to the resonator was reduced. A possible mechanism for this effect is that the thin film of liquid neon adsorbs contaminants from the surface of either the substrate or metal. These colloidal molecules are then moved during either the freezing or evaporation of the neon through thermomolecular forces that repel the molecules from the phase-change front \cite{rempelInteractionParticleAdvancing1999, mondalMaterialsPreparedFreezingInduced2022, rabaniDryingmediatedSelfassemblyNanoparticles2003}. As particles gather on the phase-change front, they can accumulate into clusters before being encapsulated, similar to effects observed in soft matter systems \cite{meijerFrozenCheeriosEffect2025}. \\

\section{\label{sec:level6}DISCUSSION AND OUTLOOK}

In this paper we have demonstrated that repeated freezing and evaporation of neon from the surface of a superconducting resonator increases the resonator's internal quality factor. Improvements of almost a factor of 2 in the single photon regime and over an order of magnitude at higher powers are observed. By analysing the temperature and power dependence of these losses and how they change over the course of the neon depositions, we conclude that these improvements in $Q_i$ are the result of decreased losses from two-level systems, opposed to a change in the energy participation ratio or from radiation-induced quasiparticles. The results are consistent with the neon processing reducing TLS losses through removal or redistribution of surface contaminants. Our conclusion is supported by AFM images of a hBN flake contaminated with polymers taken before and after neon treatment that demonstrates that residual contaminants on the surface can be moved and clumped together by crystallising and removing neon. Our results demonstrate an effective method of reducing TLS losses due to fabrication residues that could be applied to a range of superconducting or semiconducting materials that are not robust against traditional acid or plasma cleaning processes.

\section{\label{sec:level7}ACKNOWLEDGEMENTS}
We thank Dafei Jin, Xinhao Li, Erika Kawakami and Ivan Grytsenko for assistance in establishing the neon experiments in our lab. \\

A. Palmer acknowledges the support of the CSIRO Next-Gen Program. K. Matkovic and P. Russell acknowledge the support of Sydney Quantum Academy Undergraduate Summer Scholarships, and CSIRO Next-Gen Undergraduate Scholarships. L. Delventhal acknowledges a Sydney Quantum Academy Supplementary Stipend. K. Zuo acknowledges support from the ARC Center of Excellence for Engineered Quantum Systems (CE170100009),  M. Cassidy acknowledges support from a UNSW Scientia Fellowship and an Australian Research Council Discovery Early Career Research Fellowship (DE240100590). The authors acknowledge the support of the Australian Research Council Centre of Excellence for Quantum Computer Performance and Integration (CE260100009). The authors also acknowledge the facilities as well as the technical assistance of the Research and Prototype Foundry Core Research Facility at the University of Sydney, part of the NSW node of the NCRIS-enabled Australian National Fabrication Facility. The authors acknowledge the facilities and the scientific and technical assistance of Microscopy Australia at the Electron Microscope Unit (EMU) within the Mark Wainwright Analytical Centre (MWAC) at UNSW Sydney.

\bibliography{references}

\onecolumngrid

\setcounter{figure}{0}
\setcounter{section}{0}
\renewcommand{\thefigure}{S\arabic{figure}}
\renewcommand{\thesection}{\Alph{section}}

\onecolumngrid
\begin{center}
\huge\bfseries Supplementary Materials
\end{center}

\section{Resonator Design}
\label{sect:app_resdesign}
\begin{figure*}[h]
    \centering
    \includegraphics[width=0.5\linewidth]{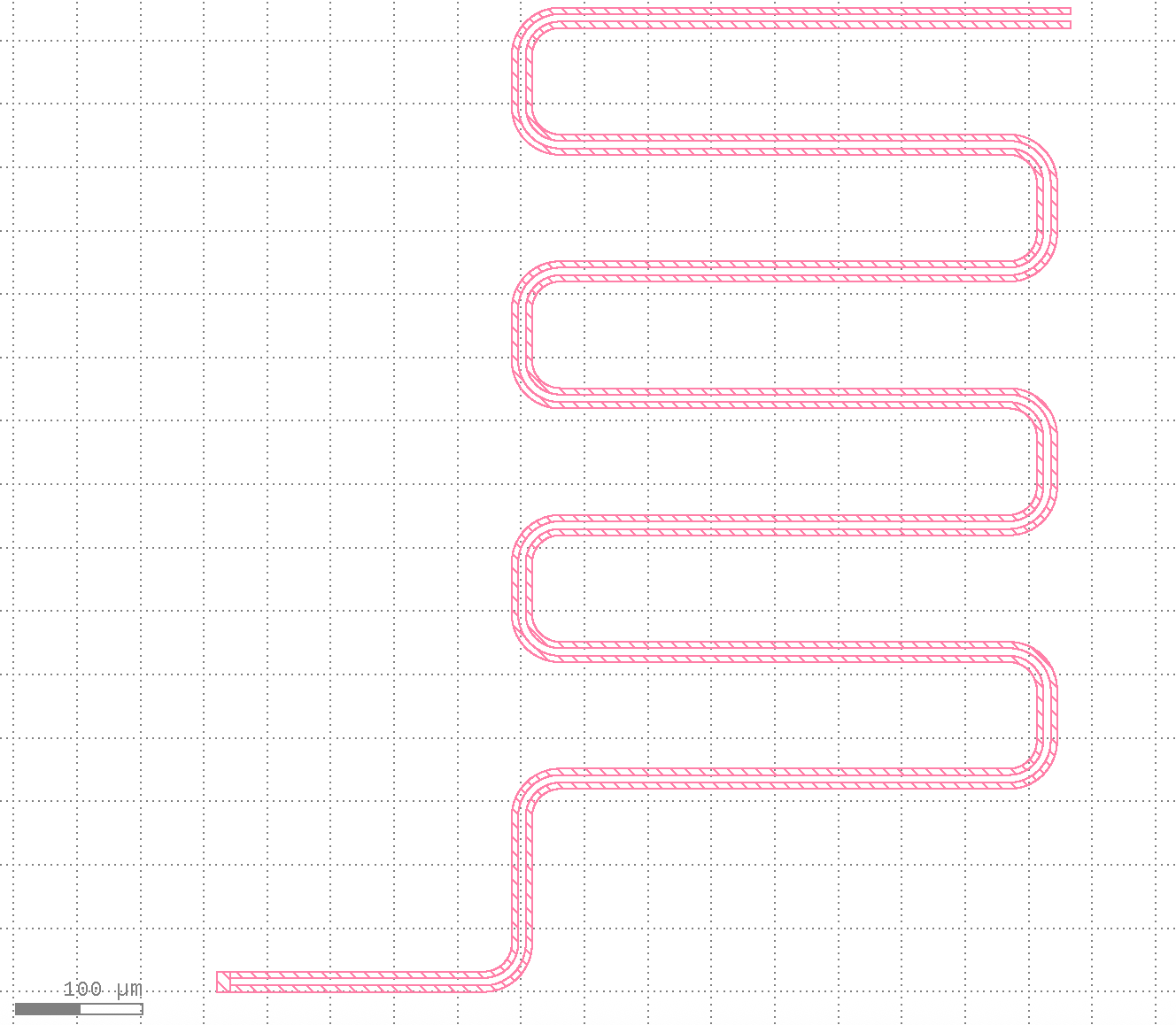}
    \caption{\label{fig:app_resdesign} Scematic diagram of resonator 1, the coloured sections indicate where NbTiN has been etched.}
\end{figure*}

The sample contains 8 resonators coupled to a single feedline with differing lengths to spread the frequencies spaced between 5.746 GHz and 6.494 GHz. Each resonator is a meandered transmission line that has been shorted on one end with the other coupled to the feedline. Resonators 1-4 have a central conductor width of 6 $\mu$m and gap 5 $\mu$m, while resonators 5-8 have a central conductor width of 10 $\mu$m and gap of 8 $\mu$m. The resonators are coupled capacitively to the feedline by a length of 200 $\mu$m and the have a meander separation of 84 $\mu$m. The feedline has a central conductor width of $15~\mu$m and a gap of $8~\mu$m. A detailed image of design for resonator 1 is shown in Figure \ref{fig:app_resdesign}.

\section{Sample Preparation}
\label{sect:app_sampleprep}

A 4 inch polished, high resistivity, silicon wafer, with a thickness of 525 $\mu$m was used as the substrate for device A. The wafer was first treated with a hydrofluoric acid (HF) dip, sputtered with a layer of 40 nm thick NbTiN and diced into 9mm $\times$ 9mm chips. The sample was then cleaned using acetone and IPA and baked at 180$^\circ$C for 5 minutes to remove any residual. The sample was coated with 200nm of ARP6200.09 e-beam resist by spin coating (4000 rpm for 60s). The resist was patterned using electron beam lithography with a Elionix ELS-F125 and developed for 60s using xylene, followed by 3:1 MIBK IPA mix for 60s and then dried with an N2 gun. The layer of NbTiN was then etched away using reactive ion etching with 4 mTor, 100W and 20 SCCM of SF6 before the e-beam resist was removed using NMP heated to 80$^\circ$C for an hour, followed by 2 minutes of $O_2$ plasma ash at 50W.\\

The patterned chips were then attached to a two-layer PCB, fabricated with Rogers 3010 with an EPIG coating, using GE-varnish which was heated to 100$^\circ$C. Two surface mount SMA connectors were soldered to the PCB and connected to the feedline input and output using 1 MIL aluminium wirebonds using a Westbond 6476D wirebonder. Wirebonds were also used to connect the ground plane of the sample to the PCB ground. The sample was packaged in a copper cell which was hermetically sealed using indium with a stainless steal tube centered above the sample for neon deposition.

\section{Measurement Setup}
\label{sect:app_measurement_setup}

\begin{figure*}[h]
    \centering
    \includegraphics[width=0.4\linewidth]{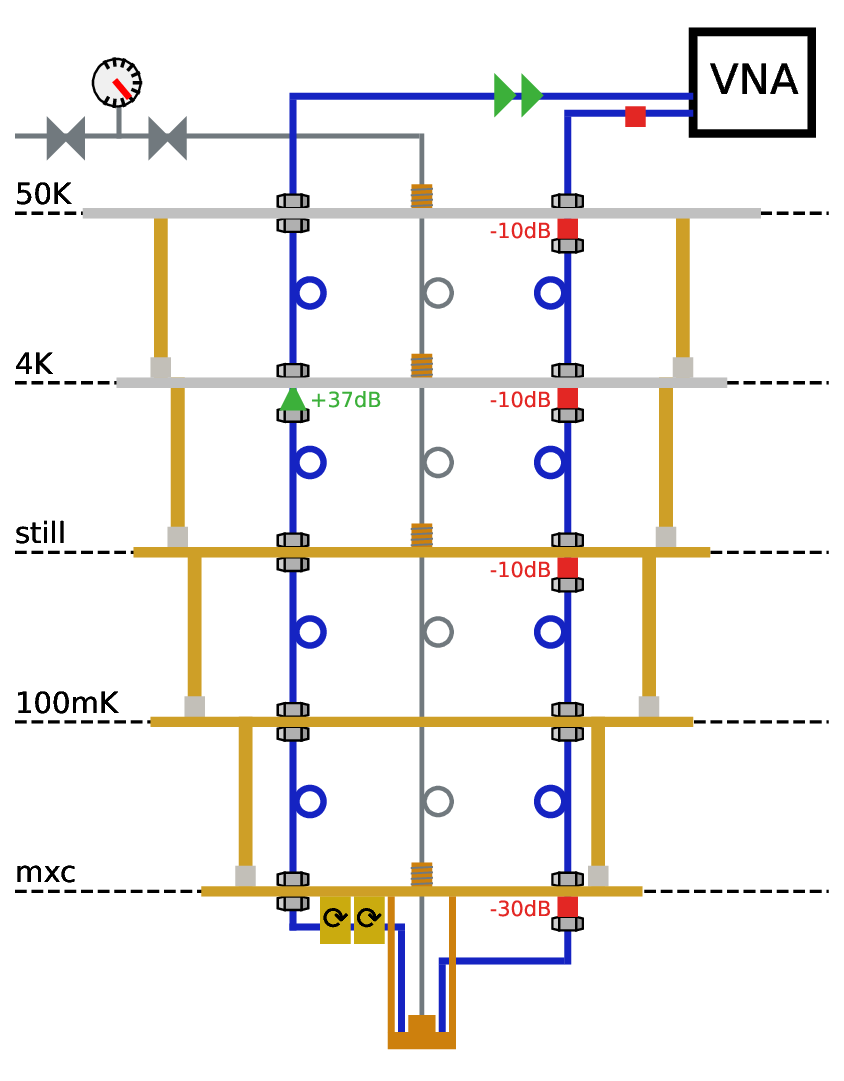}
    \caption{\label{fig:app_measurementsetup} Diagram of RF input and output setup for measurements.}
\end{figure*}

All measurements were taken in a dilution refrigerator with a base temperature of $\sim$20mK. Attenuation was placed on the input line on the 50K, 4K, still and mixing chamber (MXC) plates totalling $60$dB and a LNF-LNR1\_15A cryoamplifier with a gain of around $37$dB was added to the output line. Depending on the power level of the specific measurement, different levels of room temperature attenuation or amplification were applied. Measurements were taken using a commercial VNA at temperatures ranging from $20~\text{mK}$ to $4~\text{K}$. A diagram outlining the fridge setup shown in Fig. \ref{fig:app_measurementsetup}.

\begin{figure*}[h]
    \centering
    \includegraphics[width=0.55\linewidth]{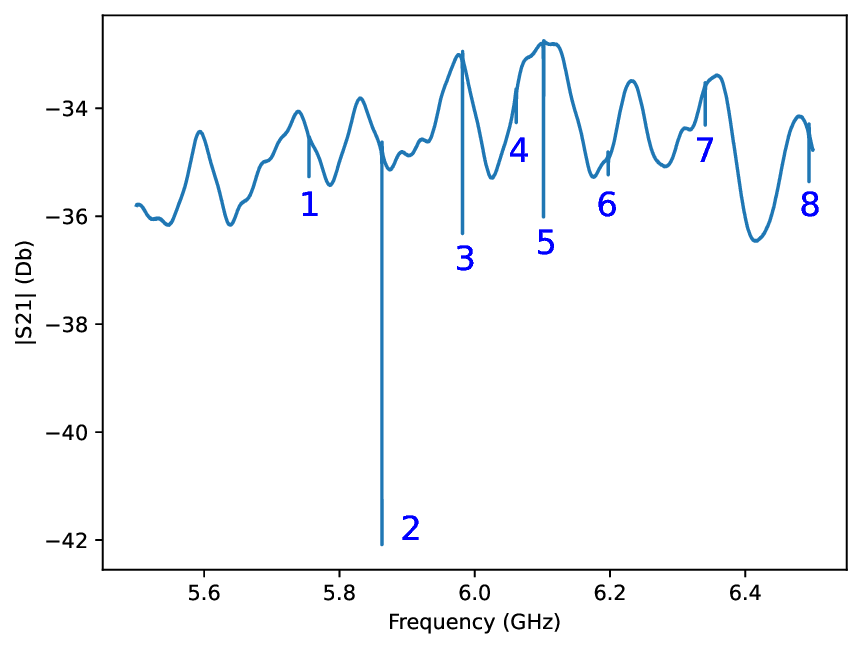}
    \caption{\label{fig:app_resonator_sweep} VNA Wide frequency sweep.}
\end{figure*}

A sweep taken across a wide frequency range showing all 8 resonances is shown in Figure \ref{fig:app_resonator_sweep}.

Table \ref{tab:Timeline} shows the timeline of the cooldowns and neon depositions. Between cooldowns the sample was removed from the fridge and removed from the sample cell.

\begin{table}[h]
    \centering
    \begin{tabular}{|c|c|}\hline
         &  Date\\\hline
         Cooldown 1&  15 September - 
16 September\\\hline
         Cooldown 2&  07 February -
12 February\\\hline
         Cooldown 2
Neon Deposition&  13 Febrary - 
15 April\\\hline
         Cooldown 3&  30 April -
20 May\\\hline
 Cooldown 3
Neon Deposition&21 May - 
02 June\\ \hline
    \end{tabular}
    \caption{Timeline of sample cooldowns.}
    \label{tab:Timeline}
\end{table}

\section{Neon Deposition}
\label{sect:app_neondep}
To coat the chip with neon, room temperature gas was pumped into the cell via a $0.18$ mm internal diameter stainless steel capillary feedline connected an external gas handling system to the cell. The gas feedline was thermally anchored to the $50~K$ and still plates of the dilution refrigerator, allowing for room temperature 5N purity neon to be deposited into the system where it crystallised on top of the chip. The amount of neon gas deposited was controlled using a small volume at a controlled pressure of $2.1$ bar between two solenoids which was repeatedly filled and then released into the sample cell which was warmed to $\sim26~K$, allowing for the neon to liquefy and flow through the tube to the sample, this setup is similar to previous neon setups \cite{zhouSingleElectronsSolid2022, matkovicCharacterizingNeonThin2025}. The presence of neon on the chip was confirmed by a shift of the resonance frequency, $f_r = \frac{1}{2\pi\sqrt{LC}}$, due to increased characteristic capacitance, $C$, as a result of the dielectric neon. \\

\section{Fitting Methodology and Results}
\label{sect:app_fits}
Complete temperature and power sweeps were taken for a subset of resonators in each cooldowns. Initially, the database contained $1,934$ frequency sweep measurements. Sweeps were excluded from analysis if the relative uncertainty in the fitted $Q_i$ exceeded $20\%$, if the temperature data was missing, or, if the temperature variance during data collection was excessively high due to control system issues, accounting for approximately $19\%$ of measurements. After excluding these data points, $1,572$ $Q_i$ measurements remained. An independent model was used for each resonator-cooldown combination. The data was fit in stages, with each stage providing the initial parameter estimates for the subsequent fitting stage.\\

To further parameterise the two-level system model shown in Equation (\ref{eq:2}), we adopt the approximations outlined in Ref. \cite{crowleyDisentanglingLossesTantalum2023}, where $T_1 \propto \tanh\left(\frac{\hbar \omega}{2k_BT}\right)$ and $(T_2)^{-1} \propto k_BT^{\beta_1}$. In addition, we introduce the fit parameter $\beta_2$ to account for the non linear response of different TLS modes to increasing number of photons and $D$ to capture the proportionality constant. This yields the final TLS model

\begin{equation} \label{eq:4}
   Q_{TLS}(\bar{n}, T) = Q_{TLS, 0} \frac{\sqrt{1 + \left(\frac{\bar{n}^{\beta_2}}{DT^{\beta_1}}\right) \tanh\left(\frac{\hbar \omega}{2k_BT}\right)}}{\tanh\left(\frac{\hbar \omega}{2k_BT}\right)}.
\end{equation}

All fits were completed using the \textit{lmfit} Python package via weighted least squares. For resonator-cooldown combinations where the temperature sweep data was not available, all measurements were taken at $20~mK$. $Q_i$ was fit using the relationship $\frac{1}{Q_i} = \frac{1}{Q_{TLS}} + \frac{1}{Q_{other}}$, assuming quasiparticle losses were negligible at low temperatures. $Q_{TLS}$ was modelled using Equation (\ref{eq:4}). The term $DT^{\beta_1}$ was combined into a single fit parameter, $D'$, due to the degeneracy of the two fit parameters, $D$ and $\beta_1$. This results in a model with four fit parameters: $D'$, $Q_{TLS, 0}$, $\beta_2$ and $Q_{other}$. The process to fit the data is outlined below:

\begin{itemize}
    \item \textbf{Model 1a}: Fit the simplified model, weighting points by the inverse of the uncertainty in the $Q_i$ estimate.
    \item \textbf{Model 1b}: If the $Q_{other}$ parameter is degenerate, refit the model with $Q_{other}$ fixed to the estimate provided in Model 1a, using the parameter fits from Model 1a as initial estimates.
\end{itemize}

Where temperature sweep data was available, the parameters from Equations (\ref{eq:4}) and (\ref{eq:5}) were fit using an iterative process that segmented the data into TLS dominated and quasiparticle dominated regimes. The model has seven fit parameters: $D$, $Q_{TLS, 0}$, $\beta_1$, $\beta_2$, $Q_{other}$, $T_c$ and $Q_{QP, 0}$. During the fit, $T_c$ was bounded between $10$ and $17~K$,  which is the expected  superconducting critical temperature for the NbTiN film. The fitting procedure is outlined below:

\begin{itemize}
    \item \textbf{Model 2a}: Fit the model, excluding quasiparticle losses, using a dataset filtered to temperatures, $T < 0.8K$, weighting points by the inverse of the uncertainty in the $Q_i$ estimate.
    \item \textbf{Model 2b}: Using the fit values from Model 2a as initial parameter estimates, fit the model on the entire dataset, fitting only quasi particle parameters, $T_c$ and $Q_{QP, 0}$. The points were weighted by the temperature to yield an initial estimate of the quasiparticle parameters that fits the high temperature regime.
    \item \textbf{Model 2c}: Fit all model parameters on the entire dataset, using the fit values from Model 2b as initial parameter estimates. The points were weighted by the inverse of the uncertainty in $Q_i$.
    \item \textbf{Model 2d}: If $Q_{other}$ is degenerate, refit the model with $Q_{other}$ fixed and using the initial estimates from Model 2c.
    \item \textbf{Final Selection}: Select the model which yields the best fit between Model 2b and 2c/2d based on the $R^2$ metric to prevent overfitting the low temperature data.
\end{itemize}

\begin{figure*}[h]
    \centering
    \includegraphics{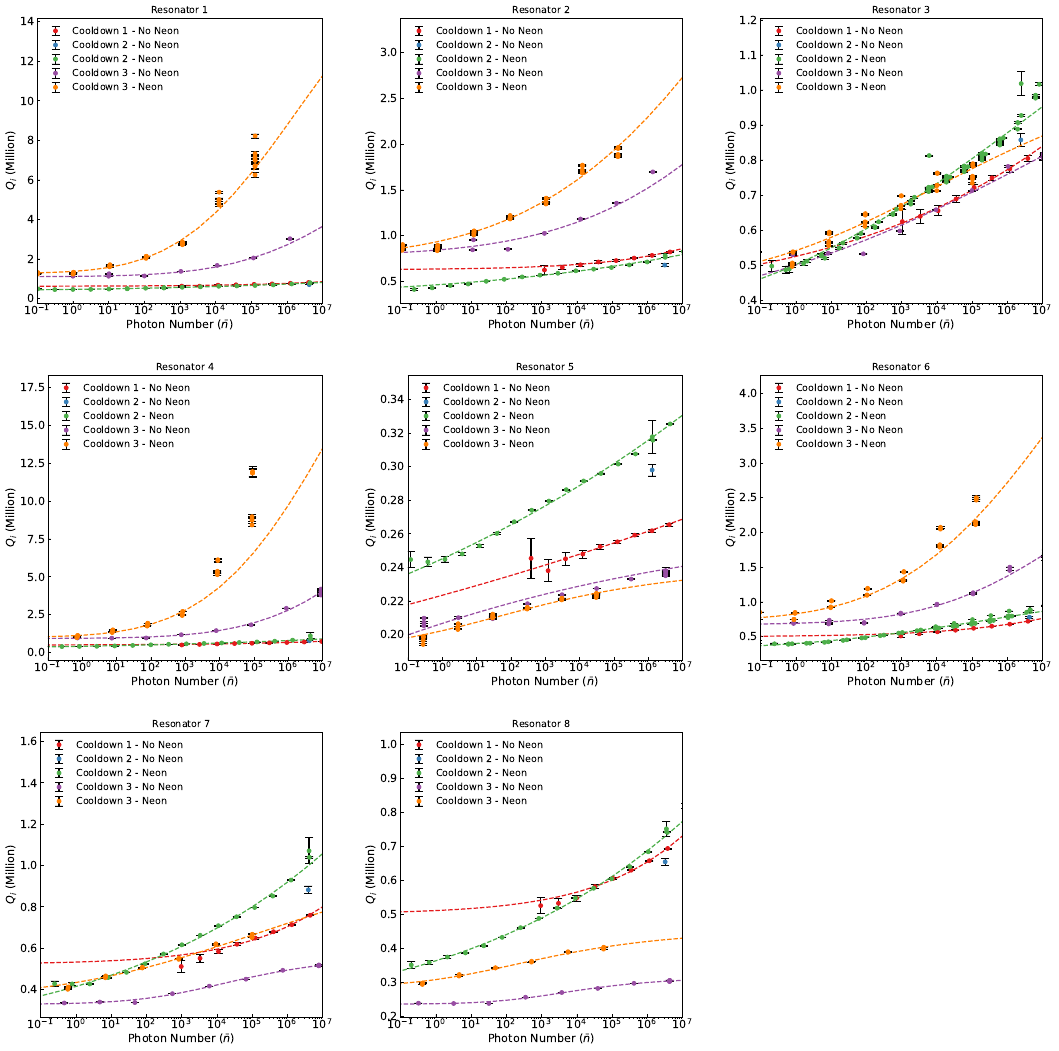}
    \caption{\label{fig:app_fits_1} $Q_i$ fits for all resonators at $20~mK$.}
\end{figure*}

Figure \ref{fig:app_fits_1} shows the fits at $20~mK$ for all resonators and cooldowns. \\

\newpage

\begin{figure*}[h]
    \centering
    \includegraphics{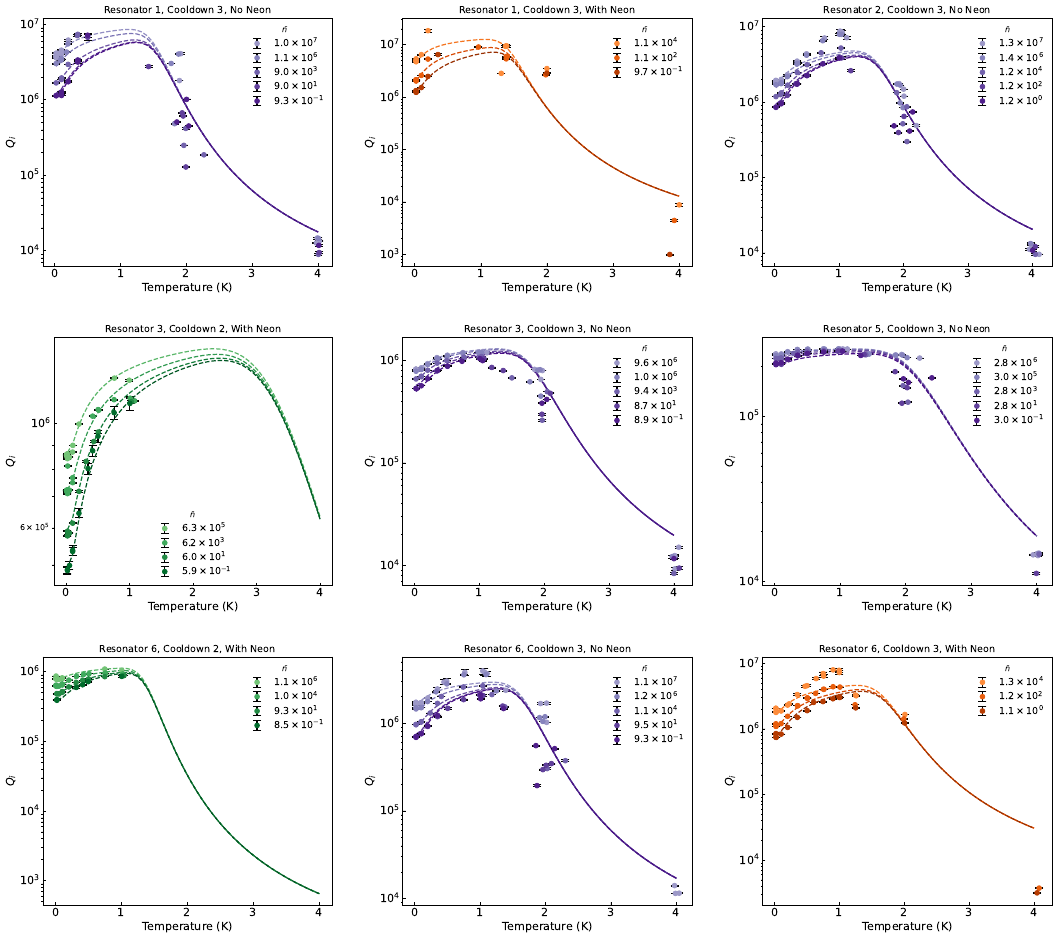}
    \caption{\label{fig:app_fits_2} Examples of $Q_i$ fits as a function of temperature}
\end{figure*}

Figure \ref{fig:app_fits_2} shows examples of the temperature and power dependent sweeps for resonators where this data was available.

\newpage

Tables \ref{tab:Fit Parameters1} - \ref{tab:Fit Parameters3} shows the fit parameters and the uncertainties for each model, where missing uncertainties have been fixed in the fitting process outlined above.

\begin{table}[h!]
    \centering
    \begin{tabular}{|c|c|c|c|c|c|c|}\hline
         Resonator & Cooldown & Neon &  Data Points & Parameter & Estimate & Rel Uncertainty\\\hline
1 & 1 & 0 & 10 & $Q_{TLS, 0}$ & 6.34 E+05 & 2.65\%\\\cline{5-7}
 &  &  &  & $D'$ & 4.29 E+01 & 40.91\%\\\cline{5-7}
 &  &  &  & $\beta_2$ & 2.27 E-01 & 8.14\%\\\cline{5-7}
 &  &  &  & $Q_{other}$ & 3.00 E+07 & \\\hline
1 & 2 & 1 & 24 & $Q_{TLS, 0}$ & 3.81 E+05 & 7.90\%\\\cline{5-7}
 &  &  &  & $D'$ & 1.85 E+00 & 36.62\%\\\cline{5-7}
 &  &  &  & $\beta_2$ & 1.21 E-01 & 9.56\%\\\cline{5-7}
 &  &  &  & $Q_{other}$ & 3.00 E+07 & \\\hline
1 & 3 & 0 & 103 & $Q_{TLS, 0}$ & 1.20 E+06 & 3.39\%\\\cline{5-7}
 &  &  &  & $D$ & 9.64 E+00 & 31.70\%\\\cline{5-7}
 &  &  &  & $\beta_1$ & -2.31 E-01 & 27.90\%\\\cline{5-7}
 &  &  &  & $\beta_2$ & 3.71 E-01 & 5.31\%\\\cline{5-7}
 &  &  &  & $Q_{other}$ & 1.33 E+07 & \\\cline{5-7}
 &  &  &  & $Tc$ & 1.00 E+01 & \\\cline{5-7}
 &  &  &  & $Q_{QP, 0}$ & 2.58 E+01 & \\\hline
1 & 3 & 1 & 86 & $Q_{TLS, 0}$ & 1.35 E+06 & 2.56\%\\\cline{5-7}
 &  &  &  & $D$ & 8.29 E-01 & 30.98\%\\\cline{5-7}
 &  &  &  & $\beta_1$ & -4.57 E-01 & 16.57\%\\\cline{5-7}
 &  &  &  & $\beta_2$ & 4.79 E-01 & 4.40\%\\\cline{5-7}
 &  &  &  & $Q_{other}$ & 1.85 E+07 & \\\cline{5-7}
 &  &  &  & $Tc$ & 1.00 E+01 & \\\cline{5-7}
 &  &  &  & $Q_{QP, 0}$ & 4.48 E+00 & \\\hline
2 & 1 & 0 & 10 & $Q_{TLS, 0}$ & 6.40 E+05 & 2.79\%\\\cline{5-7}
 &  &  &  & $D'$ & 3.96 E+01 & 42.97\%\\\cline{5-7}
 &  &  &  & $\beta_2$ & 2.22 E-01 & 8.72\%\\\cline{5-7}
 &  &  &  & $Q_{other}$ & 3.00 E+07 & \\\hline
2 & 2 & 1 & 20 & $Q_{TLS, 0}$ & 3.28 E+05 & 19.48\%\\\cline{5-7}
 &  &  &  & $D'$ & 9.57 E-01 & 68.17\%\\\cline{5-7}
 &  &  &  & $\beta_2$ & 9.93 E-02 & 14.85\%\\\cline{5-7}
 &  &  &  & $Q_{other}$ & 3.00 E+07 & \\\hline
2 & 3 & 0 & 120 & $Q_{TLS, 0}$ & 8.45 E+05 & 15.83\%\\\cline{5-7}
 &  &  &  & $D$ & 2.87 E+00 & \textcolor{red}{$>$100\%}\\\cline{5-7}
 &  &  &  & $\beta_1$ & -1.26 E-01 & 95.29\%\\\cline{5-7}
 &  &  &  & $\beta_2$ & 2.07 E-01 & 23.97\%\\\cline{5-7}
 &  &  &  & $Tc$ & 1.00 E+01 & 11.33\%\\\cline{5-7}
 &  &  &  & $Q_{QP, 0}$ & 3.04 E+01 & \textcolor{red}{$>$100\%}\\\cline{5-7}
 &  &  &  & $Q_{other}$ & 8.69 E+06 & \\\hline
2 & 3 & 1 & 44 & $Q_{TLS, 0}$ & 7.98 E+05 & 4.67\%\\\cline{5-7}
 &  &  &  & $D$ & 7.03 E+01 & \textcolor{red}{$>$100\%}\\\cline{5-7}
 &  &  &  & $\beta_1$ & 9.75 E-01 & 52.94\%\\\cline{5-7}
 &  &  &  & $\beta_2$ & 2.16 E-01 & 5.61\%\\\cline{5-7}
 &  &  &  & $Tc$ & 1.53 E+01 & \textcolor{red}{$>$100\%}\\\cline{5-7}
 &  &  &  & $Q_{QP, 0}$ & 3.27 E+01 & \textcolor{red}{$>$100\%}\\\cline{5-7}
 &  &  &  & $Q_{other}$ & 1.01 E+07 & \\\hline
3 & 1 & 0 & 10 & $Q_{TLS, 0}$ & 4.12 E+05 & 25.65\%\\\cline{5-7}
 &  &  &  & $D'$ & 1.47 E+00 & 98.08\%\\\cline{5-7}
 &  &  &  & $\beta_2$ & 9.98 E-02 & 19.83\%\\\cline{5-7}
 &  &  &  & $Q_{other}$ & 3.00 E+07 & \\\hline
3 & 2 & 1 & 174 & $Q_{TLS, 0}$ & 4.03 E+05 & 12.51\%\\\cline{5-7}
 &  &  &  & $D$ & 4.29 E-01 & 30.89\%\\\cline{5-7}
 &  &  &  & $\beta_1$ & -2.79 E-02 & 26.50\%\\\cline{5-7}
 &  &  &  & $\beta_2$ & 1.60 E-01 & 5.27\%\\\cline{5-7}
 &  &  &  & $Q_{other}$ & 1.71 E+06 & 2.94\%\\\cline{5-7}
 &  &  &  & $Tc$ & 1.60 E+01 & \\\cline{5-7}
 &  &  &  & $Q_{QP, 0}$ & 7.81 E+01 & \\\hline
\end{tabular}
    \caption{Fit model parameters.}
    \label{tab:Fit Parameters1}
\end{table}\clearpage

\begin{table}[h!]
    \centering
    \begin{tabular}{|c|c|c|c|c|c|c|}\hline
         Resonator & Cooldown & Neon &  Data Points & Parameter & Estimate & Rel Uncertainty\\\hline
3 & 3 & 0 & 111 & $Q_{TLS, 0}$ & 4.21 E+05 & 14.96\%\\\cline{5-7}
 &  &  &  & $D$ & 3.95 E-01 & 40.03\%\\\cline{5-7}
 &  &  &  & $\beta_1$ & -6.91 E-02 & 32.01\%\\\cline{5-7}
 &  &  &  & $\beta_2$ & 1.22 E-01 & 8.88\%\\\cline{5-7}
 &  &  &  & $Q_{other}$ & 1.63 E+06 & 3.62\%\\\cline{5-7}
 &  &  &  & $Tc$ & 1.00 E+01 & \\\cline{5-7}
 &  &  &  & $Q_{QP, 0}$ & 3.68 E+01 & \\\hline
3 & 3 & 1 & 44 & $Q_{TLS, 0}$ & 7.23 E+05 & 12.91\%\\\cline{5-7}
 &  &  &  & $D$ & 5.31 E-02 & \textcolor{red}{$>$100\%}\\\cline{5-7}
 &  &  &  & $\beta_1$ & -7.21 E-01 & 36.69\%\\\cline{5-7}
 &  &  &  & $\beta_2$ & 2.02 E-01 & 11.02\%\\\cline{5-7}
 &  &  &  & $Tc$ & 1.53 E+01 & \textcolor{red}{$>$100\%}\\\cline{5-7}
 &  &  &  & $Q_{QP, 0}$ & 8.57 E+01 & \textcolor{red}{$>$100\%}\\\cline{5-7}
 &  &  &  & $Q_{other}$ & 1.11 E+06 & \\\hline
4 & 1 & 0 & 10 & $Q_{TLS, 0}$ & 5.02 E+05 & 5.89\%\\\cline{5-7}
 &  &  &  & $D'$ & 1.35 E+01 & 57.05\%\\\cline{5-7}
 &  &  &  & $\beta_2$ & 1.78 E-01 & 12.73\%\\\cline{5-7}
 &  &  &  & $Q_{other}$ & 3.00 E+07 & \\\hline
4 & 2 & 1 & 19 & $Q_{TLS, 0}$ & 2.23 E+05 & 40.35\%\\\cline{5-7}
 &  &  &  & $D'$ & 3.84 E-01 & \textcolor{red}{$>$100\%}\\\cline{5-7}
 &  &  &  & $\beta_2$ & 1.14 E-01 & 12.41\%\\\cline{5-7}
 &  &  &  & $Q_{other}$ & 3.00 E+07 & \\\hline
4 & 3 & 0 & 13 & $Q_{TLS, 0}$ & 9.59 E+05 & 9.86\%\\\cline{5-7}
 &  &  &  & $D'$ & 2.31 E+01 & 59.00\%\\\cline{5-7}
 &  &  &  & $\beta_2$ & 3.89 E-01 & 16.68\%\\\cline{5-7}
 &  &  &  & $Q_{other}$ & 3.00 E+07 & 13.92\%\\\hline
4 & 3 & 1 & 39 & $Q_{TLS, 0}$ & 1.03 E+06 & 9.53\%\\\cline{5-7}
 &  &  &  & $D'$ & 3.11 E+00 & 30.00\%\\\cline{5-7}
 &  &  &  & $\beta_2$ & 4.63 E-01 & 14.90\%\\\cline{5-7}
 &  &  &  & $Q_{other}$ & 3.00 E+07 & 8.17\%\\\hline
5 & 1 & 0 & 10 & $Q_{TLS, 0}$ & 1.00 E+05 & \textcolor{red}{$>$100\%}\\\cline{5-7}
 &  &  &  & $D'$ & 2.46 E-01 & \textcolor{red}{$>$100\%}\\\cline{5-7}
 &  &  &  & $\beta_2$ & 2.77 E-02 & 93.11\%\\\cline{5-7}
 &  &  &  & $Q_{other}$ & 3.00 E+07 & \\\hline
5 & 2 & 1 & 20 & $Q_{TLS, 0}$ & 1.43 E+05 & 29.56\%\\\cline{5-7}
 &  &  &  & $D'$ & 5.01 E-01 & 87.11\%\\\cline{5-7}
 &  &  &  & $\beta_2$ & 5.03 E-02 & 19.25\%\\\cline{5-7}
 &  &  &  & $Q_{other}$ & 3.00 E+07 & \\\hline
5 & 3 & 0 & 116 & $Q_{TLS, 0}$ & 8.91 E+03 & \textcolor{red}{$>$100\%}\\\cline{5-7}
 &  &  &  & $D$ & 1.00 E-04 & \textcolor{red}{$>$100\%}\\\cline{5-7}
 &  &  &  & $\beta_1$ & 1.89 E-02 & \textcolor{red}{$>$100\%}\\\cline{5-7}
 &  &  &  & $\beta_2$ & 1.24 E-01 & 43.93\%\\\cline{5-7}
 &  &  &  & $Tc$ & 1.00 E+01 & 13.69\%\\\cline{5-7}
 &  &  &  & $Q_{QP, 0}$ & 3.09 E+01 & 50.42\%\\\cline{5-7}
 &  &  &  & $Q_{other}$ & 2.66 E+05 & 3.28\%\\\hline
5 & 3 & 1 & 38 & $Q_{TLS, 0}$ & 8.94 E+05 & 16.79\%\\\cline{5-7}
 &  &  &  & $D$ & 1.88 E-01 & \textcolor{red}{$>$100\%}\\\cline{5-7}
 &  &  &  & $\beta_1$ & -4.20 E-01 & \textcolor{red}{$>$100\%}\\\cline{5-7}
 &  &  &  & $\beta_2$ & 2.51 E-01 & 21.64\%\\\cline{5-7}
 &  &  &  & $Tc$ & 1.00 E+01 & \textcolor{red}{$>$100\%}\\\cline{5-7}
 &  &  &  & $Q_{QP, 0}$ & 1.00 E+00 & \textcolor{red}{$>$100\%}\\\cline{5-7}
 &  &  &  & $Q_{other}$ & 2.40 E+05 & 2.49\%\\\hline
    \end{tabular}
    \caption{Fit model parameters.}
    \label{tab:Fit Parameters2}
\end{table}\clearpage

\begin{table}[h!]
    \centering
    \begin{tabular}{|c|c|c|c|c|c|c|}\hline
         Resonator & Cooldown & Neon &  Data Points & Parameter & Estimate & Rel Uncertainty\\\hline
6 & 1 & 0 & 10 & $Q_{TLS, 0}$ & 5.06 E+05 & 4.11\%\\\cline{5-7}
 &  &  &  & $D'$ & 1.96 E+01 & 41.67\%\\\cline{5-7}
 &  &  &  & $\beta_2$ & 2.05 E-01 & 8.40\%\\\cline{5-7}
 &  &  &  & $Q_{other}$ & 3.00 E+07 & \\\hline
6 & 2 & 1 & 169 & $Q_{TLS, 0}$ & 3.80 E+05 & 10.73\%\\\cline{5-7}
 &  &  &  & $D$ & 8.10 E-01 & 30.52\%\\\cline{5-7}
 &  &  &  & $\beta_1$ & -2.04 E-02 & 52.53\%\\\cline{5-7}
 &  &  &  & $\beta_2$ & 2.06 E-01 & 5.58\%\\\cline{5-7}
 &  &  &  & $Q_{other}$ & 1.44 E+06 & 2.81\%\\\cline{5-7}
 &  &  &  & $Tc$ & 1.00 E+01 & \\\cline{5-7}
 &  &  &  & $Q_{QP, 0}$ & 1.00 E+00 & \\\hline
6 & 3 & 0 & 113 & $Q_{TLS, 0}$ & 8.09 E+05 & 9.36\%\\\cline{5-7}
 &  &  &  & $D$ & 8.66 E+00 & 82.52\%\\\cline{5-7}
 &  &  &  & $\beta_1$ & -1.05 E-01 & 92.56\%\\\cline{5-7}
 &  &  &  & $\beta_2$ & 3.13 E-01 & 15.37\%\\\cline{5-7}
 &  &  &  & $Tc$ & 1.00 E+01 & 13.85\%\\\cline{5-7}
 &  &  &  & $Q_{QP, 0}$ & 2.64 E+01 & \textcolor{red}{$>$100\%}\\\cline{5-7}
 &  &  &  & $Q_{other}$ & 3.93 E+06 & \\\hline
6 & 3 & 1 & 129 & $Q_{TLS, 0}$ & 7.85 E+05 & 4.87\%\\\cline{5-7}
 &  &  &  & $D$ & 1.24 E+00 & 29.25\%\\\cline{5-7}
 &  &  &  & $\beta_1$ & -1.99 E-01 & 16.94\%\\\cline{5-7}
 &  &  &  & $\beta_2$ & 3.07 E-01 & 7.25\%\\\cline{5-7}
 &  &  &  & $Q_{other}$ & 8.20 E+06 & \\\cline{5-7}
 &  &  &  & $Tc$ & 1.16 E+01 & \\\cline{5-7}
 &  &  &  & $Q_{QP, 0}$ & 1.40 E+01 & \\\hline
7 & 1 & 0 & 10 & $Q_{TLS, 0}$ & 5.32 E+05 & 4.52\%\\\cline{5-7}
 &  &  &  & $D'$ & 2.52 E+01 & 46.77\%\\\cline{5-7}
 &  &  &  & $\beta_2$ & 2.21 E-01 & 8.77\%\\\cline{5-7}
 &  &  &  & $Q_{other}$ & 3.00 E+07 & \\\hline
7 & 2 & 1 & 20 & $Q_{TLS, 0}$ & 1.81 E+05 & 52.97\%\\\cline{5-7}
 &  &  &  & $D'$ & 2.26 E-01 & \textcolor{red}{$>$100\%}\\\cline{5-7}
 &  &  &  & $\beta_2$ & 1.30 E-01 & 8.06\%\\\cline{5-7}
 &  &  &  & $Q_{other}$ & 3.00 E+07 & \\\hline
7 & 3 & 0 & 13 & $Q_{TLS, 0}$ & 7.76 E+05 & 7.23\%\\\cline{5-7}
 &  &  &  & $D'$ & 1.21 E+01 & 47.78\%\\\cline{5-7}
 &  &  &  & $\beta_2$ & 4.07 E-01 & 13.46\%\\\cline{5-7}
 &  &  &  & $Q_{other}$ & 5.68 E+05 & 2.94\%\\\hline
7 & 3 & 1 & 26 & $Q_{TLS, 0}$ & 5.21 E+05 & 7.08\%\\\cline{5-7}
 &  &  &  & $D'$ & 1.19 E+00 & 22.86\%\\\cline{5-7}
 &  &  &  & $\beta_2$ & 2.03 E-01 & 8.35\%\\\cline{5-7}
 &  &  &  & $Q_{other}$ & 1.12 E+06 & 5.54\%\\\hline
8 & 1 & 0 & 10 & $Q_{TLS, 0}$ & 5.09 E+05 & 3.30\%\\\cline{5-7}
 &  &  &  & $D'$ & 2.43 E+01 & 37.12\%\\\cline{5-7}
 &  &  &  & $\beta_2$ & 2.07 E-01 & 7.41\%\\\cline{5-7}
 &  &  &  & $Q_{other}$ & 3.00 E+07 & \\\hline
8 & 2 & 1 & 20 & $Q_{TLS, 0}$ & 2.06 E+05 & 18.86\%\\\cline{5-7}
 &  &  &  & $D'$ & 4.56 E-01 & 50.77\%\\\cline{5-7}
 &  &  &  & $\beta_2$ & 1.15 E-01 & 6.35\%\\\cline{5-7}
 &  &  &  & $Q_{other}$ & 3.00 E+07 & \\\hline
8 & 3 & 0 & 13 & $Q_{TLS, 0}$ & 9.29 E+05 & 7.66\%\\\cline{5-7}
 &  &  &  & $D'$ & 1.71 E+01 & 62.55\%\\\cline{5-7}
 &  &  &  & $\beta_2$ & 4.75 E-01 & 16.67\%\\\cline{5-7}
 &  &  &  & $Q_{other}$ & 3.15 E+05 & 1.55\%\\\hline
8 & 3 & 1 & 18 & $Q_{TLS, 0}$ & 7.53 E+05 & 7.75\%\\\cline{5-7}
 &  &  &  & $D'$ & 1.58 E+00 & 27.89\%\\\cline{5-7}
 &  &  &  & $\beta_2$ & 3.28 E-01 & 12.08\%\\\cline{5-7}
 &  &  &  & $Q_{other}$ & 4.52 E+05 & 2.28\%\\\hline
    \end{tabular}
    \caption{Fit model parameters.}
    \label{tab:Fit Parameters3}
\end{table}
\clearpage

\section{Uncertainty Estimates}
The uncertainty in the fitted value of $Q_{TLS}$ was estimated using $\sigma_{Q_{TLS}} = \sqrt{\mathbf{J}^T\Sigma\mathbf{J}}$, where $\mathbf{J}$ is the Jacobian of the $Q_{TLS}$ function and $\Sigma$ is the covariance matrix of the fitted parameters. The Jacobian, $\mathbf{J} = (\frac{\partial Q_{TLS}}{\partial Q_{TLS, 0}}, \frac{\partial Q_{TLS}}{\partial D}, \frac{\partial Q_{TLS}}{\partial \beta_1}, ...)^T$ was estimated around the point of best fit using the method of finite differences. Table \ref{tab:TLS_Uncertainty} shows  the uncertainty in the estimated $Q_{TLS}$ for the initial cooldown (cooldown 1, no neon) and the final cooldown (cooldown 3, with neon) evaluated at $T = 20~mK$ and $\bar{n} = 10^5$.

\begin{table}[h!]
    \centering
    \begin{tabular}{|c|c|r|c|r|r|}\hline
    \multirow{2}{*}{Resonator} & \multicolumn{2}{|c|}{Initial} & \multicolumn{2}{|c|}{Final} & \multirow{2}{*}{$\frac{Q_{TLS}\text{(final)}}{Q_{TLS}\text{(initial)}}$} \\\cline{2-5}
    & \rule{0pt}{3ex}$Q_{TLS}$ (Million)& $\frac{\sigma_{Q_{TLS}}}{Q_{TLS}}$ & $Q_{TLS}$ (Million)& $\frac{\sigma_{Q_{TLS}}}{Q_{TLS}}$&  \\[5pt]\hline
1 & 0.73 & 0.38\% & 9.67 & 5.54\% & 13.28 ± 0.74 \\ \hline
2 & 0.74 & 0.39\% & 2.36 & 1.81\% & 3.21 ± 0.06 \\ \hline
3 & 0.73 & 0.34\% & 2.56 & 3.63\% & 3.50 ± 0.13 \\ \hline
4 & 0.63 & 0.43\% & 8.49 & 36.07\% & 13.48 ± 4.86 \\ \hline
5 & 0.26 & 0.15\% & 3.94 & 37.51\% & 15.35 ± 5.76 \\ \hline
6 & 0.63 & 0.41\% & 2.91 & 4.48\% & 4.63 ± 0.21 \\ \hline
7 & 0.65 & 0.59\% & 1.62 & 8.31\% & 2.48 ± 0.21 \\ \hline
8 & 0.61 & 0.40\% & 4.03 & 19.99\% & 6.58 ± 1.31 \\ \hline
    \end{tabular}
    \caption{Uncertainty in $Q_{TLS}$ estimate evaluated at T = $20~mK$ and $\bar{n} = 10^5$.}
    \label{tab:TLS_Uncertainty}
\end{table}

\section{Quasiparticle Loss Impact}
\label{sect:app_quasiparticles}
\begin{figure*}[h]
    \centering
    \includegraphics{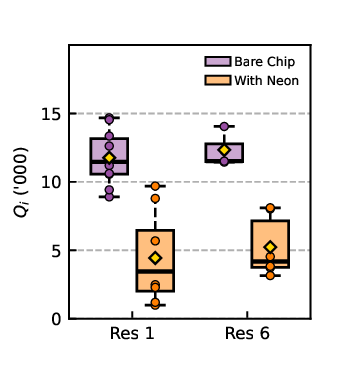}
    \caption{\label{fig:Quasiparticle_4k_comparison} Internal quality factor ($Q_i$) measured at $4~K$ for resonators 1 and 6, with and without neon. Each dot represents an individual measurement taken across various photon occupancies, and the gold diamonds indicate the mean $Q_i$ for each configuration.}
\end{figure*}

To evaluate whether the neon film alters quasiparticle induced dissipation in the device, we compare the measured internal quality factors of the resonators with and without neon at $4~K$. At this elevated temperature, thermal quasiparticle losses dominate, such that $Q_i \approx Q_{QP}(T)$. Figure~\ref{fig:Quasiparticle_4k_comparison} compares measured $Q_i$ at a number of photon occupancies for resonators 1 and 6, the resonances which could be reliably measured $4~K$. As we observe lower $Q_i$ when the neon is present compared to the bare chip, we conclude that the presence of the neon film does not reduce the quasiparticle dissipation.

\section{Device B}
\label{sect:app_device2}
A separate device was prepared using the same design, with $12~nm$ thick NbTiN film on a sapphire substrate. The sample was packaged into a sample cell and the process of neon deposition and evaporation was applied. Due to time constraints a full temperature and power characterisation was not conducted on this device. Comparing the fitted $Q_i$ from the initial and final cooldowns, this device showed improvement in $Q_i$ for 6 out of the 8 resonators on the chip. The results for each resonator are shown in Table \ref{tab:device2_results}.

\begin{table}[h!]
    \centering
    \begin{tabular}{|c|c|c|c|}\hline
    Resonator & Initial $Q_i$ ('000) & Final $Q_i$ ('000) & Relative Change (\%) \\\hline
1 & 70.8 & 150.2 & 112.3\\ \hline
2 & 79.2 & 238.0 & 200.6\\ \hline
3 & 88.9 & 213.3 & 140.0\\ \hline
4 & 153.6 & 210.5 & 37.1\\ \hline
5 & 33.9 & 98.7 & 191.0\\ \hline
6 & 8.3 & 48.4 & 481.6\\ \hline
7 & 23.6 & 16.3 & -31.0\\ \hline
8 & 272.0 & 258.7 & -4.9\\ \hline
    \end{tabular}
    \caption{Change in fitted $Q_i$ for device B evaluated at T = $20~mK$ and $\bar{n} \approx 10^5$.}
    \label{tab:device2_results}
\end{table}

\section{HBN Cleaning}
\label{sect:app_hbn}
A separate sample was prepared using the same NbTiN film as device A, patterned with a Field Effect Transistor. A flake of hexagonal boron nitride was placed on top of the patterned chip using polycarbonate tape before being cleaned with chloroform. Residual polymer particles remained on the surface as evidenced by an AFM image taken of the surface using a scanassyst-air tip, shown in Fig. \ref{fig:figure6}b. The sample was then packaged into a sample cell a thin film of neon was deposited onto the sample using the process outlined in Supplementary Materials \ref{sect:app_neondep}. The neon was then removed and the sample removed from the cell and another image was taken of the sample, shown in Fig. \ref{fig:figure6}d. This appears to show that the contaminants were clustered together and moved over micrometer scales, leaving some areas cleaned of these contaminants.

\end{document}